\documentclass[a4paper,conference]{IEEEtran}
\usepackage{algorithm}
\usepackage[noend]{algpseudocode}
\usepackage{varwidth}
\pdfoutput=1

\ifCLASSINFOpdf
  \usepackage[pdftex]{graphicx}
\else
\fi

\usepackage{amssymb}

\usepackage[switch]{lineno}

\usepackage{color}
\ifCLASSOPTIONcompsoc
    \usepackage[caption=false, font=normalsize, labelfont=sf, textfont=sf]{subfig}
\else
\usepackage[caption=false, font=footnotesize]{subfig}
\fi

\newcommand{\postfig}[0]{\vspace{-0.4cm}}
\newcommand{\sectionspace}{\vspace{-2mm}}

\begin{document}
\IEEEoverridecommandlockouts
\IEEEpubid{\makebox[\columnwidth]{\scriptsize{QoMEX2017 -- Erfurt, Germany; 978-1-5386-4024-1/17/\$31.00 \copyright2017 IEEE}}\hspace{\columnsep}\makebox[\columnwidth]{ }}

%\linenumbers %deactivate for camera-ready paper!

\title{Analysis of Problem Tokens to Rank Factors Impacting Quality in VoIP Applications}

% NOTE!: MAKE A BLIND VERSION FOR REVIEW

\author{
%\IEEEauthorblockA{for final paper use the style to the right for all authors\\
%please also make your own references blind\\
%make the acknowledgments blind \\
%do not forget to cancel it for the final paper}
%\and
%\IEEEauthorblockN{Homer Simpson}
%\IEEEauthorblockA{Twentieth Century Fox\\
%\and
\IEEEauthorblockN{Jayant Gupchup, Yasaman Hosseinkashi, Martin Ellis, Sam Johnson and Ross Cutler}
\IEEEauthorblockA{Microsoft Corporation\\
Redmond, WA, USA \\
\{jayagup, yahossei, maellis, sajohnso, rcutler\} @microsoft.com}
}


%Telephone: (800) 555--1212, Fax: (888) 555--1212}


\maketitle

\begin{abstract}
%\boldmath
User-perceived quality-of-experience (QoE) in internet telephony systems is commonly evaluated using subjective ratings computed as a Mean Opinion Score (MOS). 
In such systems, while user MOS can be tracked on an ongoing basis, it does not give insight into which factors of a call induced any perceived degradation in QoE -- it does not tell us what \emph{caused} a user to have a sub-optimal experience. 
For effective planning of product improvements, we are interested in understanding the impact of each of these degrading factors, allowing the estimation of the return (i.e., the improvement in user QoE) for a given investment.
To obtain such insights, we advocate the use of an end-of-call ``problem token questionnaire'' (PTQ) which probes the user about common call quality issues (e.g., distorted audio or frozen video) which they may have experienced. 
In this paper, we show the efficacy of this questionnaire using data gathered from over 700,000 end-of-call surveys gathered from Skype (a large commercial VoIP application). 
We present a method to rank call quality and reliability issues and address the challenge of isolating independent factors impacting the QoE. 
Finally, we present representative examples of how these problem tokens have proven to be useful in practice.
\end{abstract}

% Keywords

% \fi
\IEEEpeerreviewmaketitle

\section{Introduction}
\label{s:introduction}
\sectionspace
%This demo file is intended to serve as a ``starter file''
%for IEEE conference papers produced under \LaTeX\ using
%IEEEtran.cls version 1.7 and later.
%% You must have at least 2 lines in the paragraph with the drop letter
%% (should never be an issue)
%I wish you the best of success.

\begin{figure}[t] 
\captionsetup[subfigure]{labelformat=empty}
    \centering
  \subfloat[]{%
       \fbox{\includegraphics[width=0.89\linewidth,trim=1px 3px 1px 20px,clip]{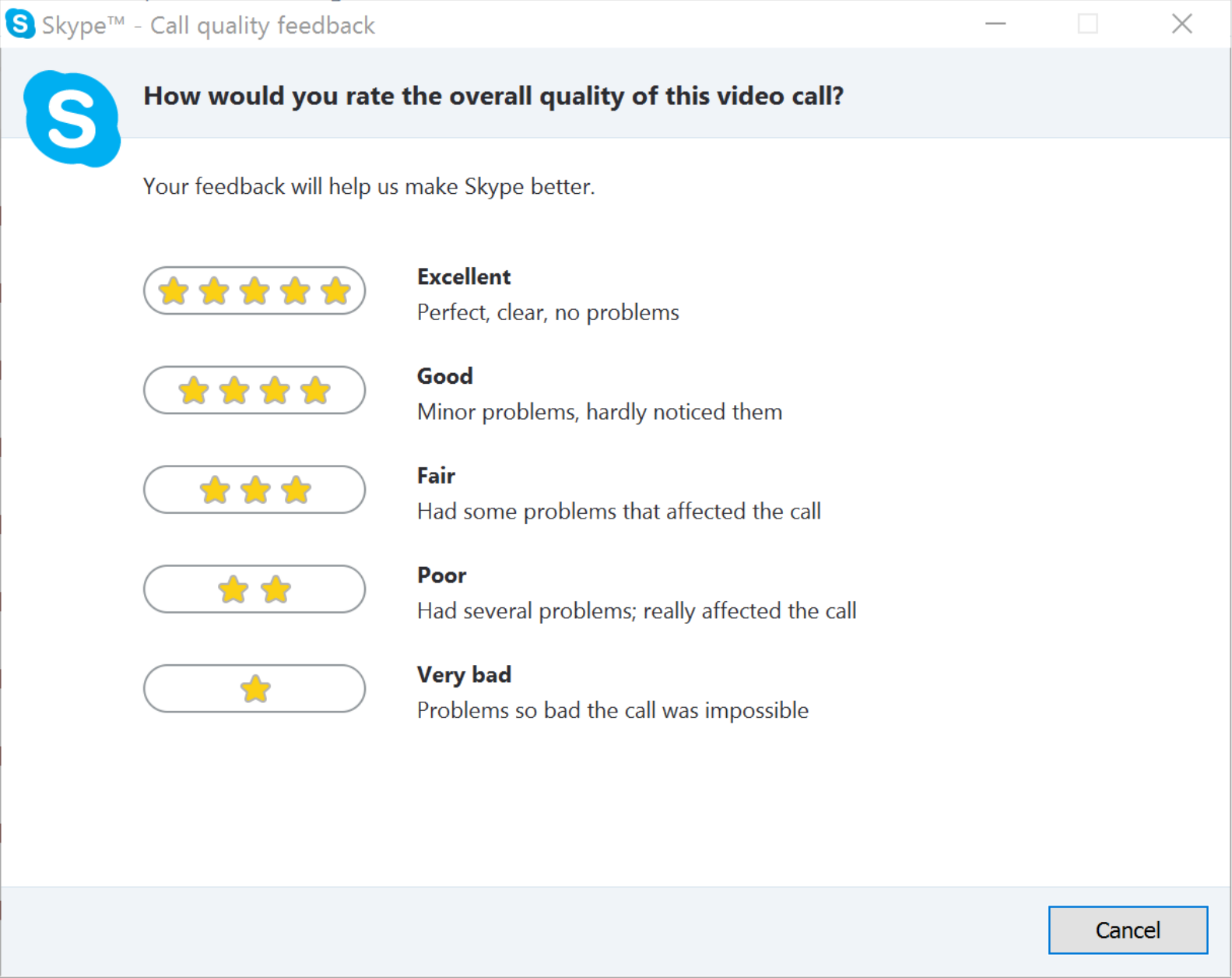}}}
%    \label{1a}
\vspace{-0.05in}
  \subfloat[]{%
        \fbox{\includegraphics[width=0.89\linewidth,trim=1px 3px 1px 20px,clip]{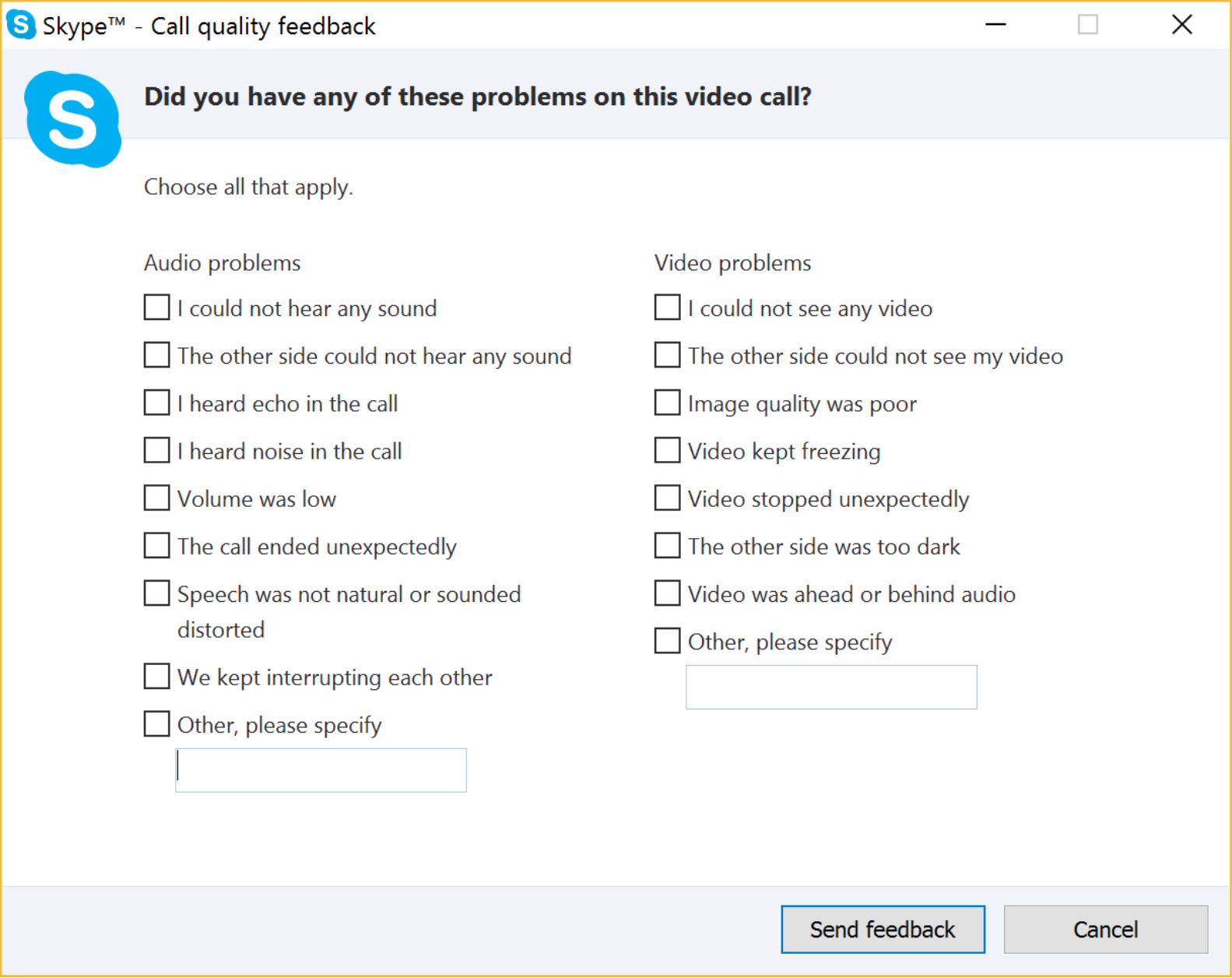}}}
 %   \label{1b}\\
  \caption{The top panel shows the Skype call quality feedback (CQF) dialog shown at the end of a call. The CQF dialog allows a user to provide an overall subjective rating. 
	The bottom panel shows the problem token questionnaire (PTQ) if the user gives an imperfect subjective rating.}
  \label{f:ptq} 
\postfig
\end{figure}

The quality of experience (QoE) of VoIP and video-based communication services is commonly reported in terms of the Mean Opinion Score (or MOS) \cite{voip-derango-2006, itut-p.800}. 
A MOS value represents an average of subjective quality scores reported by end users and ranges from $1$ to $5$ -- with $1$ being the worst quality and $5$ being perfect quality. 
While MOS ratings are useful in evaluating \emph{overall} system quality, detailed ground truth on the \emph{specific} quality degradations experienced by the user is often hard to obtain.
Therefore, in addition to prompting for the opinion score, our application presents the user with a set of follow-up options to indicate the existence of commonly experienced quality degradation which may have occurred during the call. 
We refer to these additional options as problem tokens. 
The details of the call quality feedback dialog (CQF) used to gather the opinion score, and the problem token questionnaire (PTQ) for audio and video calls is shown in Figure \ref{f:ptq}. Note that we do not present the PTQ if the user gives an opinion score of 5 -- indicating a perfect experience with ``no problems''.

% Martins table in blind version
%{\renewcommand{\arraystretch}{1.475}
%  \begin{tabular}{|p{6cm}|c|}
%  \hline
%  \textbf{Description} & \textbf{Rating}\\
%  \hline
%  Excellent -- Perfect, clear, no problems & 5 \\
%  \hline
%  Good -- Minor problems, hardly noticeable & 4 \\
%  \hline
%  Fair -- Had some problems & 3 \\
%  \hline
%  Poor -- Had several problems; really affected the call & 2 \\
%  \hline
%  Very Bad -- Problems so bad the call was impossible & 1 \\
%  \hline
%  \end{tabular}
%}

%\begin{figure}[!t]
%\footnotesize
%\centering
% \fbox{\includegraphics[width=0.89\linewidth,trim=1px 3px 1px 3px,clip]{./fig/cqf_win_classic.pdf}}
%\caption{
%The Skype call quality feedback dialog. Shown at the end of a call to allow a user to give an overall subjective rating. 
%}
%\label{f:cqf}
%\end{figure}
%
%\begin{figure}[!t]
%\footnotesize
%\centering
%{\renewcommand{\arraystretch}{1.475}
% \fbox{\includegraphics[width=0.89\linewidth,trim=1px 3px 1px 3px,clip]{./fig/problem_tokens_win_classic.pdf}}
%}
%\caption{
%The problem token questionnaire. Shown at the end of a call if the user gives an imperfect subjective rating, allowing them to indicate QoE issues in detail.
%}
%\label{f:ptq}
%\postfig
%\end{figure}

The PTQ is a rich source of data that provides us with insights into the areas where the user felt that their QoE was degraded. 
In addition to providing us information about the system quality, it also allows us to collect ground truth for improving the performance of various components. 
For example, it is extremely challenging to detect if the user experienced any ``echo'' artifacts during the call using technical statistics. 
If the system was able to reliably detect echo using technical metrics, algorithms for echo cancellation would be applied to minimize echo artifacts.
Not surprisingly, users are more likely to fill out the PTQ for calls with ratings of $1$ or $2$ (herein poor calls) compared to calls rated three or more (herein good calls); this bias in response rate is discussed further in Section \ref{s:results}. 
The share of poor calls expressed as a percentage of the total count of calls is referred to as the poor call rate ($PCR$). 
In this paper, we will use $PCR$ as the metric of choice, however the methods we present can be applied to any other VoIP quality metric, average call duration ($ACD$) being one example.

In this paper, we focus on how we use the data gathered from the PTQ to gain actionable insights. 
In the course of analysis of PTQ data, we have obtained several results that we feel are useful to the community.
Our main contributions are as follows:
\vspace{-2mm}
\begin{enumerate}
\item{We show that problem tokens are highly informative in explaining poor experiences. Problem tokens result in a 73\% reduction in entropy (information gain) of the poor call label.}
\item{We present a method to estimate the impact to quality metrics and rank of impediments as measured by problem tokens. Note that this rank significantly differs from the rank of the overall token frequencies.}
\item{We improve the estimate of $PCR$ impact on token areas by identifying factors that are relatively orthogonal using the correlation structure in the reported tokens.}
\item{We present practical applications of using problem tokens in decision making.}
\end{enumerate}

The rest of the paper is organized as follows: Section \ref{s:related-work} provides a review of the related work. 
In Section \ref{s:dataset}, we provide details of the data used for the analysis and results.
Section \ref{s:results} presents the main contributions of our work, outlining our analysis methods and the results of said analysis.
Based on our experience, Section \ref{s:discussion} discusses some practical and real-world applications of using problem tokens. 
In Section \ref{s:summary}, we summarize and outline possible future work.

\section{Related Work}
\label{s:related-work}
\sectionspace
In VoIP applications, it is common practice to correlate subjective experience ratings with telemetry gathered from the various back-end system components for evaluation.
Jiang et al.~\cite{via-jiang-2016} studied the correlation and prevalence of poor networking conditions (network jitter, packet loss, etc.) on $PCR$. 
Pessemier et al.~\cite{qoe-pessemier-2015} combined subjective quality ratings with technical metrics using a decision tree to understand the technical features that best explain the subjective ratings. 
Their study found that user-perceived quality decreases as users get more familiar with the system while the average call duration increases over a period of 120 days.
The analysis in this paper differs from the work of Pessemier et al. in the following ways: First, we correlate subjective ratings with problem data gathered from user feedback 
(as opposed to technical metrics), and second, the goal of our study is to breakdown quality metrics in terms of the rank and impact to the metric from the perspective of the user.
The decision tree approach is highly suited for troubleshooting but it does not provide a breakdown of the top-level metric into its components in an uncorrelated manner.
Moller et al.~\cite{moller2009taxonomy} outline a taxonomy structure, definitions of factors and their relationships to characterize the quality of experience.
They advocate a questionnaire framework \cite{itut-p.851} for evaluating interaction quality of experience.

There is a body of work addressing the topic of subjective quality assessment. Methods for measuring subjective audio and video quality have been defined within ITU-T Rec. P.800 \cite{itut-p.800} and P.910 \cite{itut-p.910}, 
and work continues within ITU-T's Study Group 12 to standardize new methods for objective quality assessment \cite{sg12-coverdale-2011}. 
These methods include techniques for objective measurement of audio and video quality from technical factors, such as the ITU E-model \cite{itut-g.107}, as well as full-reference metrics such as POLQA. 
These methods are generally intended for offline use, and therefore are of limited value in evaluating live systems. 
Weiss et al.~\cite{callquality-weiss-2014} evaluated different approaches to predicting the overall subjective quality of speech using the quality of individual segments of  calls. 
They found that most models (Weiss \cite{callquality-weiss-2014}, Rosenbluth \cite{rosenbluth1998itu} and ETSI models \cite{processing2006transmission}) outperformed simple averaging of MOS.

User studies have been used for decades to gather human feedback on audio/video quality for the purposes of performance evaluation. 
Traditionally, this has involved in-lab studies with a small number of participants, but more recently online crowdsourcing platforms (e.g., Amazon Mechanical Turk) have allowed sampling of wider population of users. 
This has been used for QoE evaluation in still image and audio/video scenarios \cite{qoe-chen-2009, crowdsourcing-ribiero-2011, qualitycrowd-keimel-2012}. 
By using crowdsourcing, it is possible to quickly obtain a very large number of evaluation samples, although there is additional variance in such experiments due to the lack of control compared with an in-lab study.

A number of data analysis techniques have been used to estimate the impact of predictors. 
The ideas outlined in \cite{ria-tonidandel-2011} provide a good overview of the approaches used to estimate the importance of correlated predictors.

\section{Dataset}
\label{s:dataset}
\sectionspace
The data and results reported in this paper were obtained from end-of-call surveys collected during real-world calls made using Skype.
The details of the dataset are as follows:
\vspace{-2mm}
\begin{itemize}
\setlength\itemsep{0.2em} % used to adjust spacing between list items
\item{Calls were sampled uniformly at random from users during a two week period.}
\item{Calls were one-to-one, rather than group or conference calls, and included both audio-only and video calls.}
\item{700,000 unique calls from in excess of 100,000 unique users.}
\end{itemize}

If a user rates a call less than 5 on the CQF dialog then the PTQ is shown; however, since submission is optional, some ratings do not have corresponding problem tokens.
The \emph{representative} dataset has a significantly higher percentage of calls that are labeled good calls.
For some results, we will re-sample the data at random such that the distribution of class labels (poor vs. good) is balanced. 
This secondary dataset is referred to as the \emph{balanced} dataset. Unless otherwise specified, we will report results on the \emph{representative} dataset. 
At this point, we would like to draw attention to our approach in presenting results in the rest of the paper.
Since Skype is a commercial application, we are unable to provide absolute numbers of the quality metrics.
However, we will provide relative ranks (scaled) to convey the relevant information.

\begin{figure}[t]
\begin{center}
  \includegraphics[width=\linewidth]{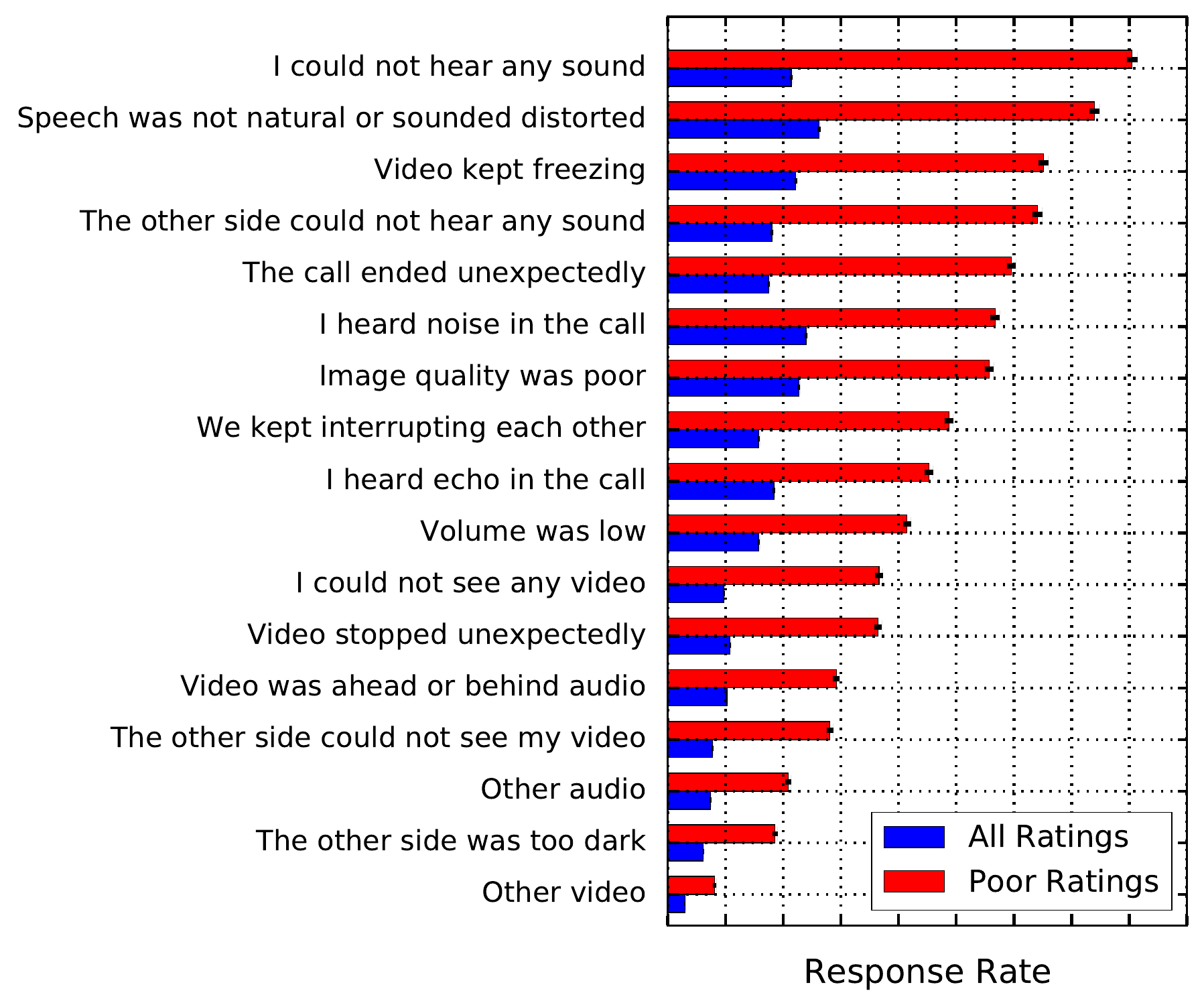}
  \vspace{-0.5cm}
  \caption{Problem token response rates for all rated calls and poor calls\protect\footnotemark.}
  \label{f:token-information}
\end{center}
\postfig
\end{figure}

\section{Analysis \& Results}
\label{s:results}

\subsection{Informativeness of Problem Tokens}
\label{ss:informativeness-of-tokens}

The percentage of the problem token selection for all rated calls and calls with poor ratings is shown in Figure \ref{f:token-information}. 
The following observations can be made based on the figure:
\vspace{-2mm}
\begin{itemize}
\setlength\itemsep{0.2em} % used to adjust spacing between list items
\item{Users are significantly more likely to respond to PTQ when the call is rated as poor compared to when the call is rated as good. For example, the response rate for ``I could not hear any sound'' token is about three
times higher for poor calls compared to the overall rated population.}
\item{The response rate sort order is different for overall rated calls and poor calls. This indicates that some problem areas are more likely to result in a poor call compared to others. }
\end{itemize}

While it is clear that users are more likely to respond to the PTQ questionnaire when they have a poor experience, the response rate (user selecting any token) is about 54\% among the poor call population, which can dilute some of the results.
In order to mitigate this bias, from here onwards our analysis considers only poor calls where the token feedback is provided. 
Note that we resample the data such that the original $PCR$ is preserved.

Computing information gain \cite{datareduction-bevington-2003} is another approach to measuring the information content present in the problem tokens. 
The information gain of two uncorrelated variables is 0.
At the other extreme, the maximum value of information gain is 1; in other words, it represents the reduction in uncertainty achieved in one variable when we know the value of the other variable. 
We compute the information gain on the balanced dataset between the $poor\_call$ indicator variable and $any\_token\_reported$ indicator variable -- a Boolean vector set to 1 if a user selected any problem token; else 0. 
The information gain for the dataset was found to be 0.73. Since this is computed on a balanced set, the information gain also represents the fractional reduction in entropy for the $poor\_call$ label if we know $any\_token\_reported$. 

%%%%%%%%%%%%%%%%%%%%%%%%%%%%%%%%%%%%%%%%
% This footnote goes with figure f:token-information
% However, this \footnotetext has to be on the same page as where the figure is displayed
\footnotetext{To preserve commercial confidentiality, absolute values are hidden in figures throughout the paper.}
%%%%%%%%%%%%%%%%%%%%%%%%%%%%%%%%%%%%%%%%

\subsection{Impact of Problem Areas on Metrics}
\label{ss:impact-on-metrics}
\sectionspace
The token frequencies (Figure \ref{f:token-information}) provide us with a ranking for prioritizing product improvement areas. 
It is worth noting that the response rate of tokens is quite different for all rated calls versus poor calls. 
For example, the percentage of users reporting  ``I could not hear any sound'' is lower than those reporting ``I heard noise in the call'' for all calls.
While the former represents a catastrophic situation where users cannot proceed with completing the desired task, the latter might be an annoyance but would not prevent completion of the desired task. 
This is the intuition behind why we see a higher rank for the token ``I could not hear any sound'' compared to ``I heard noise in the call'' when only considering poor calls.

The above intuition points to the fact that the impact to the $PCR$ metric for each problem token is related non-linearly with the overall token frequencies.
Therefore, the ranking provided by the impact to the metric is a more natural way to prioritize product improvements than considering raw token frequencies.

In order to map the token frequencies to the impact on quality metrics, we use two approaches.
The first approach relies on two assumptions:
\vspace{-2mm}
\begin{itemize}
\setlength\itemsep{0.2em} % used to adjust spacing between list items
\item{Independence: A problem token is set independently of other problem tokens.}
\item{Mutual exclusion: Users selecting a particular token would not have had a poor experience if they had not encountered this impediment. }
\end{itemize}

% ----------- TIMU procedures ------------
\algnewcommand{\LineComment}[1]{\State \(\triangleright\) #1}

\begin{algorithm}[t]
\begin{algorithmic}[1]\scriptsize

\Statex

\Procedure{TIMU}{$df, problem\_set, metric, fix\_value$}
  \State $df\_fix \gets \Call{copy}{df}$
  \State $df\_fix[problem\_set, metric] \gets fix\_value$ 
  \State $metric\_original \gets \Call{mean}{df[metric]}$ 
  \State $metric\_fix \gets \Call{mean}{df\_fix[metric]}$ 
  \State $mean\_impact \gets \Call{abs}{metric\_original - metric\_fix}$ 

  \Statex
  \LineComment{Use propogation of errors to estimate ...}
  \LineComment{uncertainty of the impact of the metric}
 
  \State $metric\_var \gets \Call{var}{df[metric]}$
  \State $metric\_fix\_var \gets \Call{var}{df\_fix[metric]}$
  \State $metric\_fix\_cov \gets \Call{covariance}{(df[metric], df\_fix[metric])}$

    %\State \begin{varwidth}[t]{\linewidth}
      % $combined\_std \gets\Call{sqrt}{metric\_var + $\par
       % \hskip\algorithmicindent $metric\_fix\_var-metric\_fix\_cov}$
      % \end{varwidth}

	\State $combined\_std \gets \sqrt{metric\_var + metric\_fix\_var - metric\_fix\_cov}$

   \Statex

%  \State $combined\_std \gets \Call{sqrt}{metric\_var+metric\_fix\_var-metric\_fix\_cov}$
%  \State $combined\_se = \frac{combined\_std}{\Call{sqrt}{\Call{rows}{df}}} $
  \State $combined\_se \gets \frac{combined\_std}{\sqrt{ df.rows }} $
  \State $mean\_impact\_95\_ci \gets 1.96 * combined\_se $

  \Statex

  \State \textbf{return} $mean\_impact, mean\_impact\_95\_ci$
\EndProcedure

\end{algorithmic}
\caption{TIMU -- Token impact on metric univariate}
\label{a:timu}
\end{algorithm}

% ----------- End of Algorithm --------------

This approach is referred to as the token impact on metric univariate (TIMU). The TIMU method is suitable for ranking impairments.
The second solution, token impact on metric multivariate (TIMM), addresses the independence and mutual exclusion assumptions. 
TIMM provides a logical grouping of problem areas, and an estimate of the impact of those areas in terms of the quality metric.
We advocate using TIMU and TIMM in conjunction -- while TIMM identifies groups and provides an estimate of the impact between groups, 
TIMU allows us to rank areas within those groups. Next, we will go into the details of the two approaches.

\begin{figure}[t]
\begin{center}
  \includegraphics[width=\linewidth]{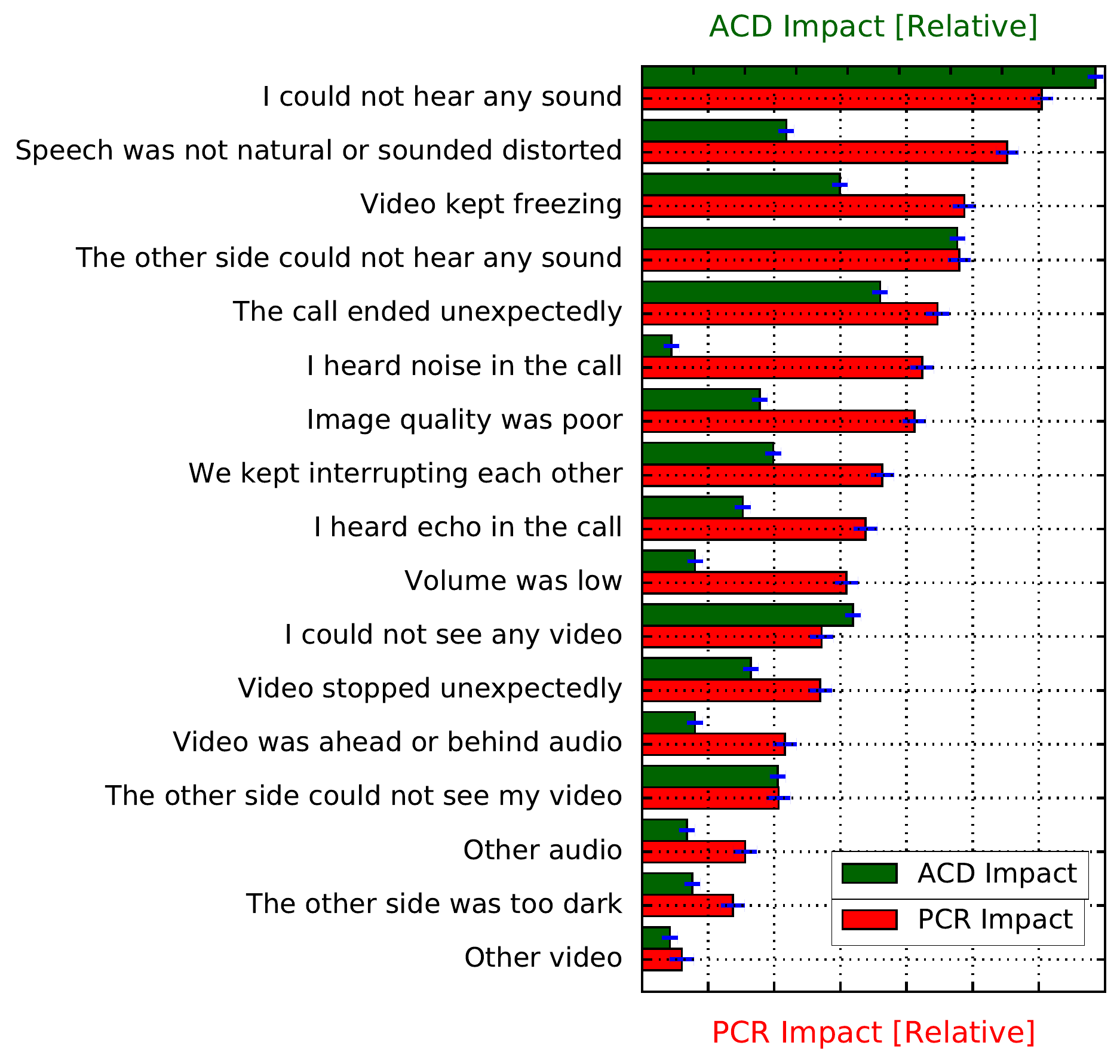}
  \caption{Estimated impact of problem tokens on PCR and ACD using TIMU.}
  \label{f:token-impact-timu}
\end{center}
\postfig
\end{figure}

\subsection{TIMU Approach}
\label{ss:timu-approach} 
\sectionspace
The TIMU approach is outlined in Algorithm \ref{a:timu}.
The idea is to estimate the impact of problem tokens on a quality metric in a univariate fashion (i.e., without considering the correlation among tokens). 
The procedure accepts the following arguments: the dataset, set of problem calls, name of the metric, and a value for the quality metric that would reflect a good experience -- for $PCR$, we pick a value of 0 indicating that call would not have been rated as poor.
For $ACD$, we pick the average of the call duration for calls where no problem is reported. 
The idea is to apply the ``fix value'' on the problem set.
The difference in the original metric and the fixed metric is the impact of the problem set on a given metric. 
Lines 8-14 show the computation of the uncertainty of the estimate using propagation of error technique \cite{datareduction-bevington-2003, errorpropagation-arras-1998}. 
This is done by combining the estimate of the variance of the original metric, the fixed metric, and the covariance among the two.

The outcome of applying the TIMU approach for $PCR$ and $ACD$ is shown in Figure \ref{f:token-impact-timu}. 
It is interesting to note that the rank of the problem areas is different for $PCR$ and $ACD$. 
The media reliability metrics (``I could not hear any sound'', ``Call ended unexpectedly'', ``I could not see any video'') have the highest impact for $ACD$. 
A number of quality areas such as ``unnatural or distorted speech'' and ``freezing video'' have more impact on $PCR$ then on $ACD$. 
We have found this approach to be very useful to rank areas that need improvement (or investment). 
One shortcoming that needs to be mentioned here is that the impact on the metric is overestimated due to the correlation and mutual exclusion assumptions. 
However, the results can be used to estimate the rank of the problem areas.
This shortcoming is addressed by the TIMM approach.

\begin{figure}[t]
\begin{center}
  \includegraphics[width=\linewidth]{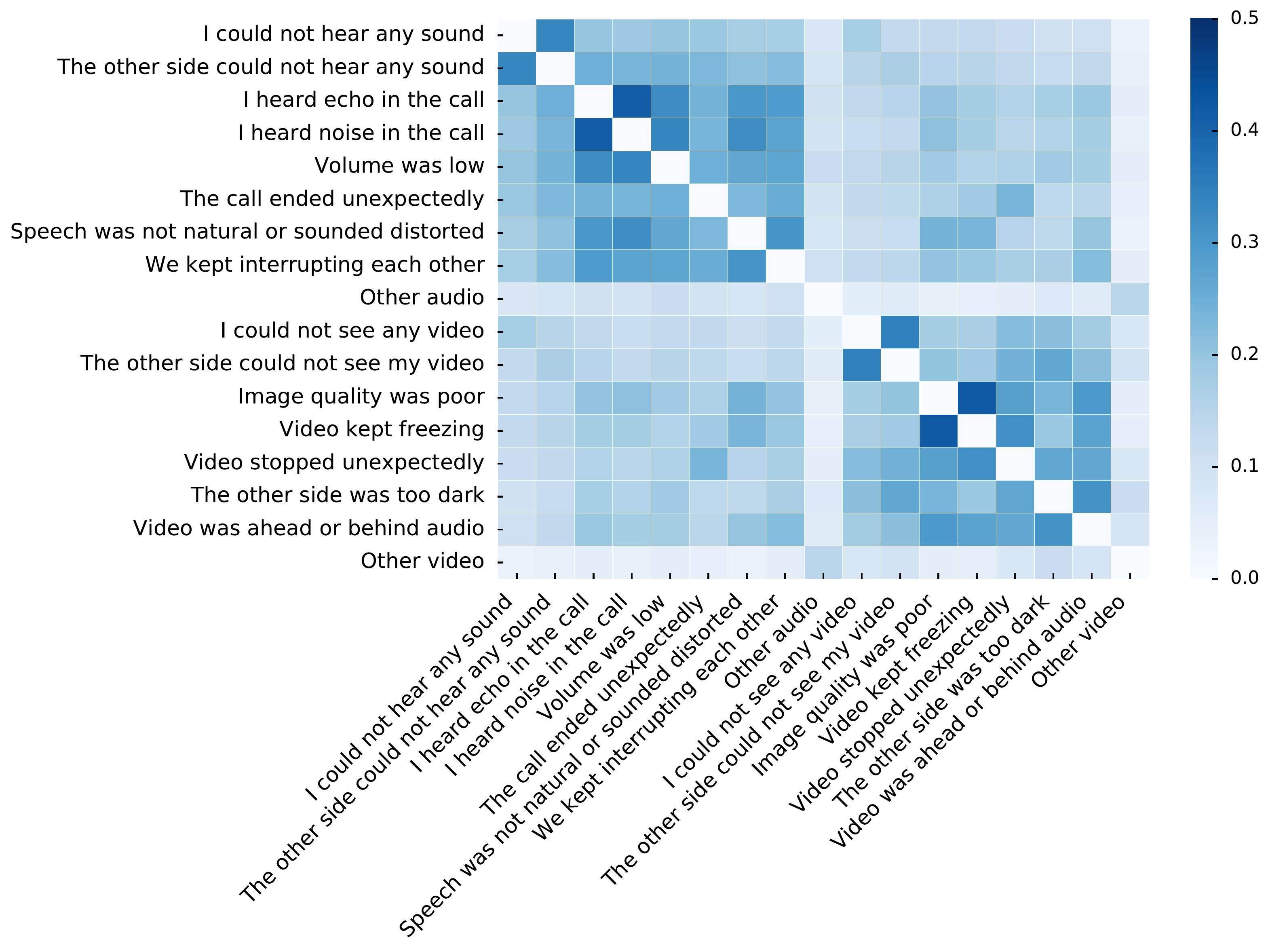}
  \caption{Jaccard similarity scores for problem tokens (diagonals set to zero).}
  \label{f:token-jacc-sim}
\end{center}
\end{figure}

Before proceeding to the TIMM approach, we further motivate the need to improve on the TIMU approach by looking at the correlation of the problem tokens.
We use the Jaccard similarity score \cite{datamining-tan-2006} to measure the degree of overlap between the tokens. 
Perfect overlap between two Boolean vectors results in a Jaccard similarity score of 1, whereas no overlap leads to a Jaccard similarity score of 0. 
The token correlations are shown in Figure \ref{f:token-jacc-sim}. 
Note that the diagonal elements have been made zero as those would always represent 1. 
We see very strong correlations among the tokens. For example, when users complain about ``echo'', more than 40\% of the time they also complain about experiencing ``noise'' in the call. 
In 34\% of cases, ``video stopped unexpectedly'' complaints are accompanied by ``poor video quality'' complaints. 
While it is our goal to make these tokens as unambiguous as possible during the design phase of these tokens (out of scope of this paper), it is clear that users perceive quality problems as a collection of problem groups rather than a single problem. 
Therefore, we need an approach that computes the impact to $PCR$ by considering these  correlations.

% ----------- TIMM procedure ------------

\begin{algorithm}[t]
\begin{algorithmic}[1]\scriptsize

\Statex

\Procedure{TIMM}{$df, metric, loading\_threshold$}
  \State \textbf{Clean Data:}
  \LineComment{Remove uninformative variables from $df$ to avoid singularity}
  \Statex
  \State \textbf{Estimate PCorr:}
  \LineComment{Estimate polychoric correlation matrix from cleaned problem token matrix}
  \Statex
  \State \textbf{Tune NumLatentFactors:}
  \LineComment{Use Parallel Analysis on PCorr to fix the number of latent factors}
  \Statex
  \State \textbf{Estimate Dimension Loadings:}
   \LineComment{Suppress weak loadings to zero using $loading\_threshold$}
   \LineComment{Generate new dimensions from the factor loading of dominant contribution}
 \Statex
  \State \textbf{Build Predictive Model for $metric$ Using Estimated Dimensions:}
  \LineComment{Fit generalized linear model (GLM)}
  \LineComment{$metricNoChange \gets $ Predict $metric$ value for mean dimension values}
  \Statex
  \State \textbf{Estimate Impact on Metric:}
  	\For{\textbf{each} $dim \in Dimension$ }
		\LineComment{$metricChange \gets $ Predict metric when the $dim$ is reduced}
		\LineComment{Use $metricChange$ and $metricNoChange$ to estimate the improvement in metric}
  	\EndFor
\EndProcedure

\end{algorithmic}
\caption{TIMM -- Token impact on metric multivariate}
\label{a:timm}
\end{algorithm}

% ----------- End of Algorithm --------------

\subsection{TIMM Approach} 
\label{ss:timm-approach}
\sectionspace
The TIMM approach is based on projecting the observed data into a lower dimensional space of meaningful factors and carrying on the estimation of impact on metrics in the lower dimensional space. This is achieved through Exploratory Factor Analysis (EFA) \cite{JohnsonWichern2007, EverittHothorn2011} and Generalized Linear Model (GLM) techniques \cite{TaylorFrancis1989}.
We skip the details of EFA and GLM methodology and instead briefly discuss the key characteristics that make these standard frameworks work so well for problem token data.
Since the problem tokens are ratings that indicate users' satisfaction (i.e., the token is set to 1 if problem is encountered and 0 if not encountered), these tokens can be modeled as dichotomous observations of a continuous trait, say ``satisfaction level''.
If satisfaction level dips lower than a certain threshold, the user rates 1, otherwise 0.
This way, the observed variable is binomial while the latent variable is continuous. 
The correlation structure between latent continuous variables is estimated from binary observations using the Polychoric correlation coefficient \cite{Pearson1900, Olsson1979}. 
We compared EFA on Polychoric correlation to Principal Components Analysis (PCA) on Pearson correlation.
In addition to the theoretical incompetence of Pearson correlation coefficient for binomial data, this approach does not preserve class separability (i.e., separation between good and poor calls), nor provided interpretable results. %We abandoned this line of investigation.
An overview of the TIMM procedure is provided in Algorithm \ref{a:timm}.

The Polychoric correlation coefficient proved to be highly effective in revealing meaningful groupings of problem tokens through EFA (with varimax rotation \cite{EverittHothorn2011}). 
A 5-dimensional subspace of rotated factors with dominant loadings accounts for 81\% of total variability in the 15-dimensional space of problem tokens. These factors are not orthogonal
as in PCA \cite{JohnsonWichern2007}, but provide a reasonable trade-off between interpretability and dimensionality reduction.
The weak remaining correlation between factors is captured in the GLM model through interaction effect terms. 
By dropping the tokens with small loading from each factor (we used a threshold of 0.5), the problem groups (PGs) shown in Table \ref{t:problem-token-factors} are uncovered. 

\begin{table}[!t]
\caption{Factors extracted from problem tokens}
\label{t:problem-token-factors}
\centering
\renewcommand{\arraystretch}{1.2}
\begin{tabular}{|l|l|}
\hline
\textbf{Problem Groups} & \textbf{Problem Tokens}\\
\textbf{(\%Variance explained)}        & \\
\hline
Audio Quality (26\%) & We kept interrupting each other\\
                     & Speech was not natural or sounded distorted \\
                     & Volume was low \\
                     & I heard echo in the call \\
                     & I heard noise in the call \\
\hline
Video Quality (25\%) & The other side was too dark\\                               
                     & Video stopped unexpectedly\\
                     & Video was ahead or behind audio\\
                     & Image quality is poor\\
                     & Video kept freezing\\
\hline
One-way Video (12\%) & I could not see any video\\
                     & The other side could not see my video\\
\hline
One-way Audio (11\%) & I could not see any sound\\
                     & The other side could not see my sound\\
\hline
Reliability (7\%)    & The call ended unexpectedly\\
\hline
\end{tabular}
\end{table}

Logistic regression is used to predict the reduction in $PCR$ by fixing each of the problem groups shown in Table \ref{t:problem-token-factors}.
The most accurate model consists of all the main effect terms (the PGs) and two interaction effect terms; specifically between two pairs of PGs: Audio Quality (PG1) and Video Quality (PG2), and Audio Quality (PG1) and One-way Audio (PG4).
In practical terms, this means that when PG1 and PG2 (and similarly PG1 and PG4) are reported \emph{together}, they have an impact different to the sum of their individual contributions.
%Model detail is shown in Equation \ref{LR}. 
%\begin{eqnarray}\label{LR}
%PCR = &-&3 + 8.5 PG1 + 8.8 PG2 + 7.4 PG3 \\ \nonumber
%&+& 8.5 PG4 + 3.4 PG5 - 16.3 PG1\times PG2  \\ \nonumber
%&-&  13.3 PG1  \times PG4 + \epsilon
%\end{eqnarray}
The Area Under Curve (AUC) using this approach is 95\%; this is a significant improvement over the baseline approach of using $any\_token\_reported$. 
The baseline method has a false positive rate (FPR) of $10.8\%$ and a true positive rate (TPR) of $48\%$. At the same FPR, the logistic regression model
has a TPR of $93\%$ resulting in a significant improvement in performance. 

\begin{figure}[t!]
\begin{center}
  \includegraphics[width=\linewidth]{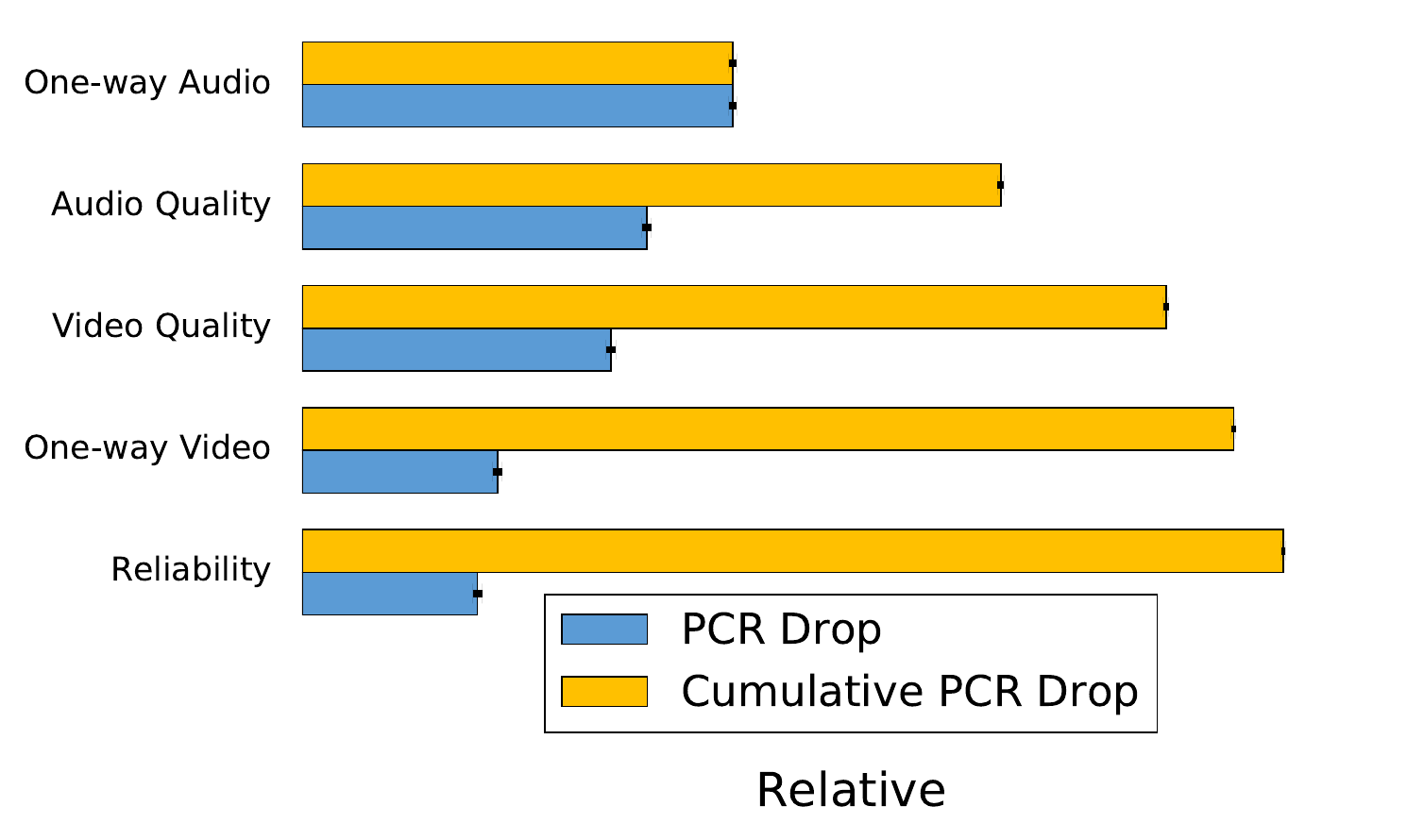}
  \caption{Predicted maximum relative reduction in $PCR$ using TIMM.}
  \label{f:predicted-pcr-reduction}
\end{center}
\postfig
\end{figure}

Figure \ref{f:predicted-pcr-reduction} shows the maximum reduction that can be achieved by fixing a single PG at a time. 
The blue bars indicate the reduction in $PCR$ if a single PG is fixed while all other PGs still occur at their current level.
In the population we studied, the data indicates that fixing One-way Audio has the highest return on investment (RoI) while Reliability shows the smallest RoI in terms of user satisfaction.
This provides the priority in problem groups and helps formulate efforts to fix them within our study population.
Note that the values shown in blue are not additive since they represent the drop in $PCR$ assuming only one problem group is fixed. However, the interaction terms in the model help to predict the combined effect. 
Yellow bars in Figure \ref{f:predicted-pcr-reduction} demonstrate the expected cumulative drop in $PCR$. It is worth mentioning that we see TIMU and TIMM methods providing complementary information.
While TIMM provides an estimate of RoI in fixing problem groups, TIMU provides a relative ranking withing the problem group.

\section{Discussion}
\label{s:discussion}
\sectionspace
In our experience, problem tokens have served as a useful source of data in solving many practical decision making challenges. Here, we outline some representative examples.

\subsection{Analysis of quality for new releases/versions}
\label{ss:release-analysis}

When new versions of Skype are released, engineering teams are keen to track the user-perceived QoE.
This is usually done by comparing the quality metrics of the new release to previous releases; typically, regressions in quality attract more attention than improvements.
Upon discovering that a quality metric has regressed, the natural response is to ask which changes in the product have caused this regression. However, this is not always an easy question to answer.
A typical release contains a number of changes that can interact with each other in complex ways. 
These changes may not be detected in component, integration or end-to-end regression tests, but once released may interact under certain hardware or network conditions previously unknown -- resulting in poor experiences for potentially millions of users.
On numerous occasions, we have used the problem token data as a first response to reduce the search space of the quality regression. 
For example, one release contained a change to bandwidth allocation logic, a corner case resulted in a sharp uptick in $PCR$ and the response rate of the ``I could not hear any sound'' token. This allowed us to narrow down the underlying problem resulting in a faster turnaround time for the fix.

\subsection{Unbiased comparisons when updating system components}
\label{ss:unbiased-comparisons}
Problem token data has been useful in evaluating the user experience when making systemic changes in components. 
The problem in evaluating systemic changes is that the technical metrics are often not comparable between the two systems. 
For example, when making a major overhaul in the jitter control component, we were unable to use the technical metrics to compare the two systems, since the definitions of the metrics themselves had changed. 
However, the associated problem tokens (``We kept interrupting each other'', ``Speech was not natural or sounded distorted'') are based on user feedback, and can therefore be used to compare the two components.

\section{Summary}
\label{s:summary}
\sectionspace
In this paper, we analyze the value of the end-of-call ``problem token questionnaire'' in Skype calls.
Using a dataset collected from over $700,000$ calls, we show that problem tokens give useful insights in understanding the areas where our users perceive a quality degradation. 
We show that instead of relying on the raw token frequencies of problem tokens, these data can be used more effectively by estimating the impact on quality metrics. 
Towards this goal, two approaches are presented with the requirement that results are easy to interpret and take action on.

The TIMU method is used to rank the problem areas that are impacting quality metrics experienced by users.
The TIMM method exploits the correlation structure of the problem tokens to learn categories, and estimates of impact to the quality metrics within those categories. 
The goal of these two methods is to provide the next level of detail by breaking down a quality metric, this is then primarily used to estimate areas that require improvement.
We also share some practical examples of how problem tokens can be employed by engineering teams for effective decision-making in situations where technical metrics are not easily available. 

We note that the design of the PTQ (as with any questionnaire) is a key factor for the effectiveness and response rate of these tokens.
Techniques for effective design include keeping the question set small, using clear and unambiguous text, and randomizing presentation order to minimize priming bias \cite{krosnick2010question}. 
However, we defer discussion of these issues to future work.

To conclude, we would like to emphasize that understanding the overall impact of the problem tokens provides us with a very natural way to measure user-perceived QoE, and has allowed us to make investments to improve it.

\section*{Acknowledgments}
\label{s:acknowledgments}
We would like to thank Mu Han, Robert Aichner, and the Skype call quality data science team for useful discussions on problem token analysis.

%\emph{(Acknowledgments hidden for blind review)}
%Reserve space to acknowledge the people and teams that helped us in collecting this data.



%\bibliographystyle{IEEEtran}
%\bibliography{problem-tokens}

\end{document}